\documentclass[aps,superscriptaddress,superbib,twocolumn,groupedaddress]{revtex4}

\usepackage[highlight]{neel}

\bibpunct{}{}{,}{s}{}{}

\graphicspath{{fig/}}

\usepackage[T1]{fontenc}

\def\figurename{Fig.}

\usepackage[paperwidth=8.5in, paperheight=11in]{geometry}
\usepackage{graphics} 
\usepackage{times}
\usepackage{bbm}
\usepackage{tabularx}
\usepackage{tabulary}
\usepackage{tabu}
\usepackage{sistyle}
\usepackage{url}
\usepackage{float}
\usepackage{pdfpages} 

\usepackage{physics}

\newcommand{\CoNi}[2]{Co\textsubscript{$#1$}Ni\textsubscript{$#2$}}

\begin{document}

\title{Time-resolved imaging of \OE rsted field induced magnetization dynamics in cylindrical magnetic nanowires}

\author{M.~Sch\"{o}bitz}
\email[Author to whom correspondence should be addressed: ]{michael.schobitz@cea.fr}
\affiliation{Univ.\ Grenoble Alpes, CNRS, CEA, Spintec, 38054 Grenoble, France}
\affiliation{Friedrich-Alexander Univ.\ Erlangen-N\"{u}rnberg, Inorganic Chemistry, 91058 Erlangen, Germany}
\affiliation{Univ.\ Grenoble Alpes, CNRS, Institut N\'{e}el, 38042 Grenoble, France}
\author{S.~Finizio}
\affiliation{Swiss Light Source, Paul Scherrer Institut, 5232 Villigen PSI, Switzerland}
\author{A.~De Riz}
\affiliation{Univ.\ Grenoble Alpes, CNRS, CEA, Spintec, 38054 Grenoble, France}
\author{J.~Hurst}
\affiliation{Univ.\ Grenoble Alpes, CNRS, CEA, Spintec, 38054 Grenoble, France}
\author{C.~Thirion}
\affiliation{Univ.\ Grenoble Alpes, CNRS, Institut N\'{e}el, 38042 Grenoble, France}
\author{D.~Gusakova}
\affiliation{Univ.\ Grenoble Alpes, CNRS, CEA, Spintec, 38054 Grenoble, France}
\author{J.-C.~Toussaint}
\affiliation{Univ.\ Grenoble Alpes, CNRS, Institut N\'{e}el, 38042 Grenoble, France}
\author{J.~Bachmann}
\affiliation{Friedrich-Alexander Univ.\ Erlangen-N\"{u}rnberg, Inorganic Chemistry, 91058 Erlangen, Germany}
\affiliation{Institute of Chemistry, Saint-Petersburg State Univ., 198504 St.\ Petersburg, Russia}
\author{J.~Raabe}
\affiliation{Swiss Light Source, Paul Scherrer Institut, 5232 Villigen PSI, Switzerland}
\author{O.~Fruchart}
\email[Author to whom correspondence should be addressed: ]{olivier.fruchart@cea.fr}
\affiliation{Univ.\ Grenoble Alpes, CNRS, CEA, Spintec, 38054 Grenoble, France}

\date{\today}

\begin{abstract}
Recent studies in three dimensional spintronics propose that the \OE rsted field plays a significant role in cylindrical nanowires. However, there is no direct report of its impact on magnetic textures. Here, we use time-resolved scanning transmission X-ray microscopy to image the dynamic response of magnetization in cylindrical \CoNi{30}{70} nanowires subjected to nanosecond \OE rsted field pulses. We observe the tilting of longitudinally magnetized domains towards the azimuthal \OE rsted field direction and create a robust model to reproduce the differential magnetic contrasts and extract the angle of tilt. Further, we report the compression and expansion, or breathing, of a Bloch-point domain wall that occurs when weak pulses with opposite sign are applied. We expect that this work lays the foundation for and provides an incentive to further studying complex and fascinating magnetization dynamics in nanowires, especially the predicted ultra-fast domain wall motion and associated spin wave emissions.
\end{abstract}

\maketitle


Continuous developments in time-resolved magnetic imaging techniques have allowed for a shift of interest from systems that lend themselves more readily to imaging, such as flat nanostrips\cite{bib-LIT2016,bib-JUG2019,bib-FIN2020b}, to more intricate systems such as three dimensional (3D) nanostructures, with added complexity from the volume\cite{bib-DON2020,bib-DON2020b}. Such 3D nanostructures can now be fabricated with increasing ease\cite{bib-BOC2017,bib-WIL2017b,bib-SKO2020}, making the exploration of intriguing predicted magnetic configurations\cite{bib-FRU2017b}, such as domain walls (DWs) in M\"{o}bius strips or hopfions\cite{bib-GRY2020}, feasible. 
A textbook case for such an investigation is provided by cylindrical magnetic nanowires (NWs), featuring a special type of DW, the Bloch-point wall (BPW)\cite{bib-FRU2014}, which exhibits an azimuthal curling of magnetic moments around a Bloch-point on the NW axis\cite{bib-FEL1965,bib-DOE1968}. 
The dynamics of these walls are not yet well understood, but stand out compared to DWs in flat nanostrips due to fascinating theoretical predictions of fast, stable speeds\cite{bib-WIE2010}, accompanied by the controlled emission of spin waves\cite{bib-HER2016}.
Recent experiments in NWs have shown that the \OE rsted field induced by nanosecond current pulses plays a key role in stabilising walls exclusively of the BPW type, and further, imposes an azimuthal circulation parallel to the field\cite{bib-FRU2019b}. This allows controlling the wall structure and enables fast DW motion with speeds $>\SI{600}{m/s}$ with an absence of Walker breakdown. 
The \OE rsted field induced BPW circulation switching was further studied in a simulation and theory work, revealing a complex mechanism of the switching process, involving nucleation and annihilation of pairs of vortex and anti-vortex\cite{bib-FRU2021}. Further, magnetic moments in longitudinally magnetized domains are predicted to align azimuthally with the \OE rsted field, with the degree of tilt related to a competition between magnetic exchange and Zeeman energy. Although some works have investigated the \OE rsted field in nanowires\cite{bib-OTA2015,bib-AUR2013,bib-FER2020}, its major influence on magnetization dynamics in these systems has become clear rather recently and is especially pertinent in a low current density regime where the effect of spin-transfer torque is negligible\cite{bib-FRU2021b}. However, few of the predictions have been confirmed experimentally.

\begin{figure*}
\centering
\includegraphics{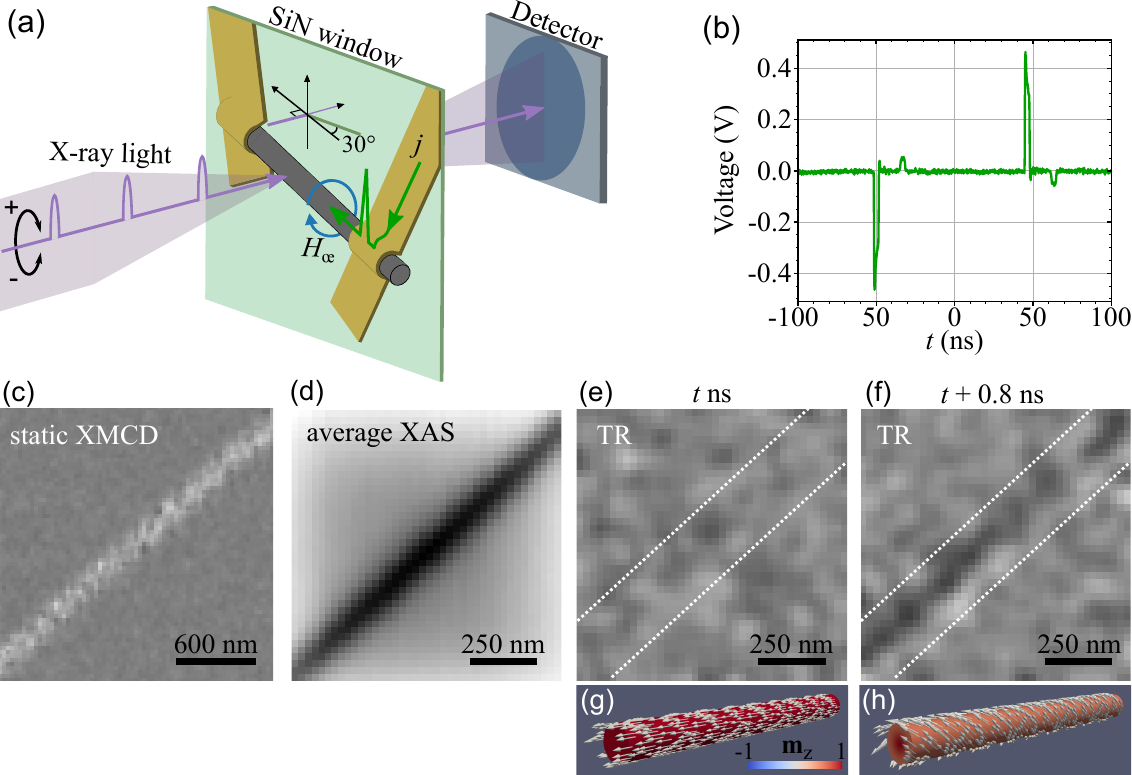}
\caption{Time-resolved STXM with electrically contacted magnetic NWs. a) Schematic of the STXM set-up, with incoming circularly-polarized X-ray photon bunches (purple) incident at \SI{30}{\degree} to the normal of the SiN window. In time-resolved mode, the frequency of the current pulses (green) inducing the \OE rsted field (blue) is phase-locked with the photon bunch frequency. b) Voltage pulse signal measured after the NW across a \SI{50}{\Omega} load, with $(+)$ and $(-)$ pulses of \SI{3}{ns} applied to induce a current density of $j=\SI{1.1\times{10^{12}}}{A/m^2}$ in a \SI{93}{nm} diameter NW. c) static XMCD image of a longitudinal domain in a \SI{93}{nm} diameter NW. d) Average of all XAS frames in one time-resolved image series equivalent to  $I_\mathrm{S}$ in eq.~\eqref{eq:TheoryTR}. e,f) Frames from a time-resolved image series showing the differential magnetic contrast observed in a \SI{93}{nm} diameter \CoNi{30}{70} NW with the wire edges indicated by guides to the eye. No current is flowing at time $t$ (e), but at time $t+\SI{0.8}{ns}$ (f) the \SI{3}{ns} current pulse with amplitude $\SI{1.1\times{10^{12}}}{A/m^2}$ is being applied. g,h) Illustrations of the magnetization in the NW at the time corresponding to frames (e,f).}
\label{fig:TRframes}
\end{figure*}

Here, we make use of time-resolved scanning transmission X-ray microscopy (STXM) to image magnetization dynamics in NWs subjected to nanosecond pulses of \OE rsted field. 
Magnetically-soft \CoNi{30}{70} NWs with diameters of $93$, $97$ and \SI{101}{nm} were electrodeposited in anodized alumina templates and freed by dissolution of the template\cite{bib-BOC2017}. Wires were dispersed onto \SI{200}{nm} thick, $100\SI{\times100}{\micro m^2}$ wide X-ray transparent Si$_3$N$_4$ windows, suspended in a $\SI{5\times5}{mm^2}$ intrinsic Si frame. Individual wires were lithographically contacted with Au pads to allow for the injection of nanosecond pulses of electric current, in turn creating an \OE rsted field around the NW, as in \subfigref{fig:TRframes}{a}. Magnetic images were acquired using STXM at the PolLux bending magnet beamline at the Swiss Light Source\cite{bib-RAA2008}. The sample was tilted by $\SI{30}{\degree}$ with respect to the X-ray beam direction and aligned so that the NW was oriented to be as parallel as possible to the beam direction. Optical sensitivity to magnetization was achieved due to the X-ray magnetic circular dichroism (XMCD) effect, whereby circularly-polarized X-ray light is absorbed differently depending on whether the magnetization is (anti-)parallel to the photon propagation direction. 
Magnetization dynamics were observed with time-resolved STXM as shown in \subfigref{fig:TRframes}{a}, making use of the intrinsically pulsed nature of synchrotron radiation (purple) and phase locking their frequency with the excitation signal (\ie current pulses in green)\cite{bib-FIN2018}. 
Time-resolved image series comprised of 1021 frames, each spaced by \SI{200}{ps}, were acquired stroboscopically with a temporal resolution of \SI{70}{ps} and spatial resolution $\approx\!\SI{40}{nm}$. Each frame is a X-ray absorption spectroscopy (XAS) image acquired with circularly-polarized light, with the magnetic contribution to intensity superimposed on the spectroscopy image. Differential magnetic contrast was revealed by a division of each XAS frame by the average of all XAS frames, the latter essentially being a XAS image of the static magnetic state (full videos in supplementary material).

To set ideas, consider the example of a time-resolved series with two frames where dynamics with opposite magnetic changes cause a variation in intensity of $\pm I_\mathrm{D}$. The intensity in each frame is given as $I_\mathrm{S}\pm I_\mathrm{D}$, with $I_\mathrm{S}$ the static contribution to the transmitted intensity. The average of the two frames is simply the static part, $I_\mathrm{S}$, and thus the differential magnetic contrast in each frame is given by
\begin{equation}
(I_\mathrm{S}\pm I_\mathrm{D}) / (I_\mathrm{S}) = 1\pm I_\mathrm{D}/I_\mathrm{S}
\label{eq:TheoryTR}
\end{equation}
While static XMCD contrast (usually of the order of a few percent) arises from the difference between XAS images taken with opposite polarizations of light, differential magnetic contrast is much weaker and of the order of $0.1\%$. 
It should be noted that if the dynamics do not lead to $\pm I_\mathrm{D}$ that is symmetric about zero, great care must be taken in the analysis, as discussed later on.

We first investigated the effect of the \OE rsted field on uniformly-magnetized domains in \CoNi{30}{70} NWs. Suitable regions with $>\!\!\SI{5}{\micro m}$-long domains were detected using static XMCD STXM\bracketsubfigref{fig:TRframes}{c}, after which a $<\!\!\SI{1}{\micro m}$ section was chosen within this. 
A repeating signal of \SI{3}{ns} alternating positive $(+)$ and negative $(-)$ voltage pulses with amplitude $\SI{1.1\times{10^{12}}}{A/m^2}$ was applied to a \SI{93}{nm} diameter NW\bracketsubfigref{fig:TRframes}{b}. Pulses were spaced by \SI{100}{ns} to allow for sufficient heat dissipation.
The frames displayed in \subfigref{fig:TRframes}{e,f} show snapshots of the differential magnetic contrast observed before and during the application of the $(+)$ current pulse. 
Before the application of current in (e), no contrast is observed since magnetization is at rest along the NW long axis (see the corresponding illustration in \subfigref{fig:TRframes}{g}) as it is for the majority of frames in the time-resolved series.
During the application of the \SI{3}{ns} $(+)$ current pulse (f), a bipolar contrast is observed across the NW, indicating the tilting of magnetization to become more parallel (black) or antiparallel (white) to the X-ray beam direction. The wire magnetization is thus tilting towards the wire azimuthal direction (see illustration \subfigref{fig:TRframes}{h}), consistent with the direction of the \OE rsted field.
Once the pulse has ended, the magnetization returns to its relaxed state and the differential magnetic contrast is no longer observed. The second pulse, with opposite polarity, gives rise to an inverted contrast.

\begin{figure}
\centering
\includegraphics{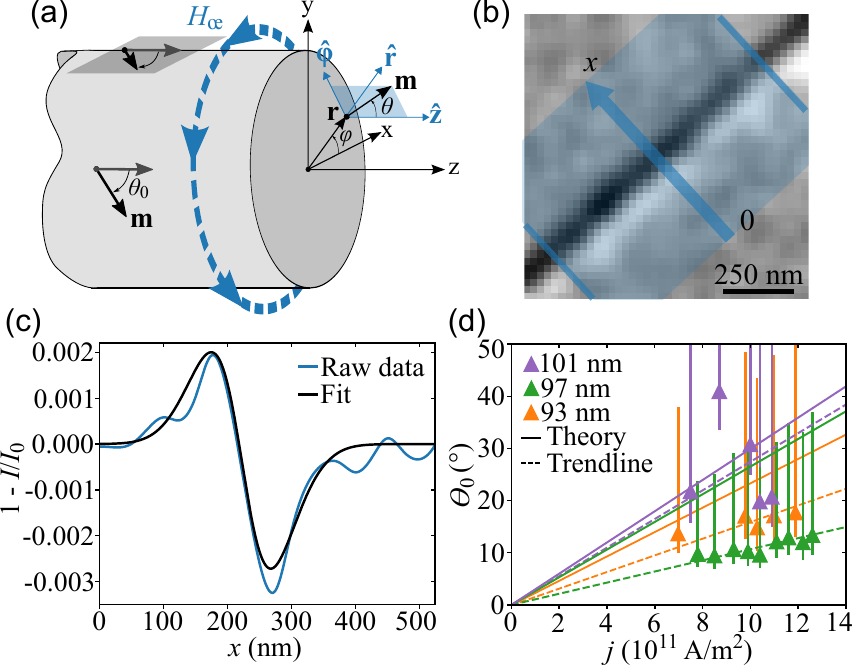}
\caption{Tilting of magnetization in NW domains as a function of current density. a) Schematic of a NW with magnetic moment, $\vect{m}$, described with spherical polar angles $\theta$ and $\phi$ in the cylindrical coordinate system ($r$,$\varphi$,$z$). An applied \OE rsted field tilts $\vect{m}$ from $\hat{\vect{z}}$ towards $\hat{\boldsymbol{\varphi}}$ by an angle, $\theta$. b) Average differential magnetic contrast of all 15 frames acquired during the application of a \SI{3}{ns} current pulse with amplitude $\SI{1.1\times{10^{12}}}{A/m^2}$ in a \SI{93}{nm} diameter NW. The width-averaged line scan is indicated by the blue arrow. The corresponding intensity profile is shown in blue in (c). The black curve is the fit from our model, with $\theta_0=\SI{18.9}{\degree}$. d) $\theta_0$ as a function of applied current density, $j$, in three different NW samples with diameters $93$, $97$ and \SI{101}{nm}, and with vertical error bars. Trendlines through each data-set and and analytical solutions (see eq.~\eqref{eq:Theory}) for each wire diameter are shown as dashed and solid lines, respectively.}
\label{fig:Theta}
\end{figure}

The signal-to-noise ratio is increased by taking the average of the 15 frames acquired during the \SI{3}{ns} current pulse\bracketsubfigref{fig:Theta}{b}.
A width-averaged line scan across the wire (blue area in \subfigref{fig:Theta}{b}) reveals the profile of this bipolar contrast (blue in \subfigref{fig:Theta}{c}), with asymmetric peak amplitudes $0.2$ and \SI{-0.3}{\%}. This is expected to be a signature of the tilting of magnetization due to the \OE rsted field, where appropriate fitting may extract quantitative information. In the following, we propose a model to reproduce this relatively simple physical situation and hence fit the recorded contrast profile to estimate the tilt angle. Our model considers the  radius-dependent degree of tilt within a NW cross section, the absorptivity of X-rays in the material and the X-ray beam spot size, which make up the key components of magnetic transmission X-ray imaging\cite{bib-DON2020}.

We use an Ansatz to describe the tilt by an angle $\theta$ of magnetic moment, $\vect{m}$, towards the azimuthal direction when subjected to the \OE rsted field\bracketsubfigref{fig:Theta}{a}:
\begin{equation}
\theta \left(r\right) = \theta_0 \sin{\left(\frac{\pi}{2}\frac{r}{R}\right)}
\label{eq:Ansatz}
\end{equation}
$\vect{r}$ is any point within the circular wire cross section, $r = \abs{\vect{r}}$, $R$ is the wire radius and $\theta_0 = \theta \left( r = R\right)$. Comparisons with micromagnetic simulations showed that this Ansatz accurately describes the tilt within the wire\cite{bib-FRU2021}.

In magnetic materials, the absorptivity, $\mu$, or linear rate of absorption of X-rays, depends on the chemical composition, the X-ray energy and the magnetization direction versus the polarity of circularly-polarized X-ray light. For a \SI{100}{\%} circularly $(+)$ or $(-)$ polarized X-ray beam parallel to the magnetization, the absorptivity is given as $\mu_{\pm}$ and we additionally define the average absorptivity, $\mu_{\mathrm{av}}=\left( \mu_-+\mu_+\right)/2 $ and the difference in absorptivity, $\Delta\mu =\mu_- -\mu_+ $. The amplitude of the latter is of particular importance, as it directly influences the strength of the XMCD effect and thus relates to the strength of any differential magnetic contrast.
Values for $\mu_{\mathrm{av}}$ and $\Delta\mu$ of the studied \CoNi{30}{70} NWs at the Co L3 absorption edge can be extracted from XAS images acquired with circularly-polarized light in static STXM. The reduced X-ray intensity behind a NW is described by the Beer-Lambert law, linking the exponential decay of light intensity through matter with $\mu$. Intensity profiles taken across the NW can thus be fitted with the law, however, a non-zero X-ray spot size must be accounted for by a convolution with a Gaussian with width $\sigma$. A detailed analysis is shown in the supplementary material. 
The fitting procedure then provides the only free parameters, absorptivity, $\mu$, and spot size, $\sigma$. The analysis relies on the $\SI{30}{\degree}$ angle between the NW magnetization and the X-ray beam, induced by the sample holder orientation. It is also due to this angle that a geometrical adjustment must be made to calculate $\Delta\mu$ from the extracted $\mu$ (see supplementary material).

Using this fitting procedure on multiple XAS images, we determined $\mu_{\mathrm{av}} = \SI{0.006\pm{0.002}}{nm^{-1}}$ and also found an average $\sigma = \SI{50\pm{4}}{nm}$, where the uncertainty is the standard deviation of the data sets. For comparison, the theoretical absorptivity for \CoNi{30}{70} at the Co L3 edge is $\mu_\mathrm{av,th} = \SI{0.019}{nm^{-1}}$, which is a factor 3 larger than our extracted values\cite{bib-NAK1999}. This is a common feature in STXM imaging, partly related to background intensity incident on the X-ray detector. We expect this amounts to $\approx\!\SI{33}{\%}$ at the PolLux beamline STXM and arises from higher-order light from the monochromating mirror ($\approx\!\SI{15}{\%}$)\cite{bib-FLE2007} and leakage of the zone plate center stop ($\lesssim\SI{18}{\%}$). Accounting for this via a subtraction from the static XAS images, we correct $\mu_{\mathrm{av}} = \SI{0.011\pm{0.005}}{nm^{-1}}$ which is closer to the theoretical value. The reason for the remaining difference is unclear. 
We similarly correct $\Delta\mu$ from $\SI{0.002\pm{0.001}}{nm^{-1}}$ to $\SI{0.003\pm{0.002}}{nm^{-1}}$ (the uncertainty also being the standard deviation). Accounting for the $\approx\!\SI{50}{\%}$ degree of circular polarization of X-ray light in this experiment, we find a $\Delta\mu = 0.006 \pm \SI{0.004}{nm^{-1}}$. This can be compared to the theoretical value, which again for the Co L3 edge is $\Delta\mu_\mathrm{th}  = \SI{0.01}{nm^{-1}}$\cite{bib-NAK1999}. Our calculated value is again lower than theory, but still in reasonable agreement considering the uncertainty. 
The background subtraction was therefore applied to all acquired STXM images. It should be noted that considering the derived values of $\mu_{\mathrm{av}}$ and $\Delta_\mu$ as effective, allows extracting the exact magnetization direction if the values were calibrated on uniform magnetization at $\SI{30}{\degree}$.

To now fit the bipolar contrast profile in \subfigref{fig:Theta}{c}, we re-use the Beer-Lambert law and include non-uniform magnetization, such as described by our Ansatz in \eqref{eq:Ansatz}:
\begin{equation}
I\left(x\right) = I_0 \exp{ - \int{ \left(    \mu_{\mathrm{av}} +\frac{1}{2}\Delta\mu\; \hat{\vect{k}}\cdot\vect{m}    \right) \diff{\ell}  }  }
\label{eq:TR-intensity}
\end{equation}
This describes the progressive absorption of X-rays through each elementary segment with length, $\diff{\ell}$, and with $\hat{\vect{k}}\cdot\vect{m}$ the component of magnetization along the X-ray beam direction, $\hat{\vect{k}}$. $I$ depends on $\vect{m}\left(r\right)$ which itself depends on $\theta_0$. 

The intensity profile is then convoluted with the Gaussian to account for the finite spot size, already determined from the static XAS image analysis. By performing the same image calculation as in \eqref{eq:TheoryTR} with the dynamic ($\theta_0 \neq 0$) and the static ($\theta_0 = 0$) intensity profile, the bipolar differential magnetic contrast profiles can be reproduced. The only free parameter for the fit is $\theta_0$, while all other variables are fixed as previously determined from the XAS image analysis. We revisit now the magnetic image in \subfigref{fig:Theta}{b} and the corresponding contrast profile in blue in \subfigref{fig:Theta}{c}, which is fit very well ($r^2=0.94$) by the black curve. The model also reproduces the asymmetric signal, which originates in part from the non-linear change in $\hat{\vect{k}}\cdot\vect{m}$ due to the geometry of the \SI{30}{\degree} sample holder and in part from the non-linear X-ray absorption due to the exponential nature of the Beer-Lambert law. From the fit we determine the tilt of magnetic moments on the surface as $\theta_0=\SI{18.9}{\degree}$, caused by the current-pulse-induced \OE rsted field.
This value comes with multiple sources of uncertainty for the magnitude of the extracted $\theta_0$. First, we determined that the uncertainty in the NW diameter measured from scanning electron microscopy images and the uncertainty in the spot size extracted from the XAS analysis translate to an uncertainty of $\lesssim\! \SI{10}{\%}$ in $\theta_0$.
This remains moderate compared with the last source, the uncertainty in $\mu_{\mathrm{av}}$ and $\Delta\mu$, for which the impact is critical due to the aforementioned exponential in \eqref{eq:TR-intensity}. We repeat the fitting of the data in \subfigref{fig:Theta}{c} using values for $\mu_{\mathrm{av}}$ of $0.006$ and \SI{0.016}{nm^{-1}} and $\Delta\mu$ of $0.001$ and \SI{0.004}{nm^{-1}}, which correspond to an uncertainty of one standard deviation (see supplementary material Fig.~S5). This gives $\theta_{0,\mathrm{max}}=\SI{47.8}{\degree}$ and $\theta_{0,\mathrm{min}}=\SI{13.2}{\degree}$, which is strongly asymmetric about $\theta_0=\SI{18.9}{\degree}$ due to the exponential in \eqref{eq:TR-intensity}. These values of $\theta_{0,\mathrm{max}}$ and $\theta_{0,\mathrm{min}}$ hence provide the range of uncertainty that we expect for our extracted value of $\theta_{0}$. However, for the case of the intensity profile produced with $\theta_{0,\mathrm{max}}$, the fit to the data is quite poor ($r^2=0.83$) indicating that the original fit with $\theta_0=\SI{18.9}{\degree}$ is more appropriate.

This fitting analysis to determine $\theta_0$ was applied to multiple time-resolved image series from three wire diameters and several applied current densities, $j$, with the results plotted in \subfigref{fig:Theta}{d}. The vertical error bars for each data point represent the range of uncertainty calculated as described above. The most significant result is that $\theta_0$ increases linearly with $j$, as also shown by the dashed trendline (constrained to pass through the origin) for each data set. While the linear dependence of the $93$ and \SI{97}{nm} diameter wires is clear, the \SI{101}{nm} diameter wire exhibits a larger spread. The reason most likely being a variation in the imaging conditions from one time-resolved series to another (when settings were changed), resulting in a change in the incident X-ray intensity. This directly impacts the signal-to-noise ratio in the differential images and possibly the applied background subtraction, as the latter may not scale linearly with the incident light intensity. In both instances the result would be a vertical offset (of unknown direction) of the data. Imaging conditions (settings) were kept constant during the acquisition of each of the other two sample data sets, however, were adjusted between the samples. For any $j$, $\theta_0$ increases from the $97$ to $93$ to \SI{101}{nm} diameter wire and using the trendline, we find a tilt rate in $\theta_0$ equivalent to $10.7$, $16.0$ and $\SI{27.4}{\degree}$ per $\SI{10^{12}}{A/m^2}$, respectively.

To compare the experimental $\theta_0$ we use an analytical model developed by A. De Riz \textit{et al}\cite{bib-FRU2021}, balancing the competition between exchange and Zeeman \OE rsted energies to describe $\theta_0\left(j\right)$ in longitudinal domains in NWs. To first order, this reads:
\begin{equation}
\theta_0 \approx \frac{1}{3.16}\frac{\mu_0 j \Ms R^3}{\pi A}
\label{eq:Theory}
\end{equation}
with $\mu_0=4\pi\times10^{-7}$ the vacuum magnetic permeability, $\Ms$ the spontaneous magnetization of the material and $A$ the exchange stiffness. De Riz \etal found that for $j<\SI{3\times{10^{12}}}{A/m^2}$, the model is an accurate description and matches well with simulations, meaning that it is appropriate to compare to the results presented here. Using magnetic parameters for \CoNi{30}{70} NWs ($\Ms=\SI{0.77}{MA/m}$\cite{bib-JIL1988} and $A=\SI{1.5\times{10^{-11}}}{J/m}$\cite{bib-TAL2002}), \eqref{eq:Theory} is plotted as solid line for each NW diameter in \subfigref{fig:Theta}{d}. The theory confirms the experimentally observed linear dependence of $\theta_0$ on $j$, however, predicts larger tilt rates. These are $\theta_0=23.4$, $26.5$ and \SI{29.9}{\degree} per $\SI{10^{12}}{A/m^2}$ for $93$, $97$ and \SI{101}{nm} diameter wires, respectively. Even though there is $\lesssim\!\SI{55}{\%}$ discrepancy between the theory and experiment, the results are promising as they indicate that the analysis is appropriate within its range of uncertainty. The experiment fails to reproduce the theoretically predicted $1/R^3$ dependence, however, we expect this to be linked to systematic errors such as the changes in imaging conditions discussed previously. Further, local inhomogeneities in the nanowire (\eg small changes in diameter or crystal grains) can also lead to a systematic offset of the data both along $\theta_0$ or $j$ which cannot be accounted for.

\begin{figure}
\centering
\includegraphics[width=\linewidth]{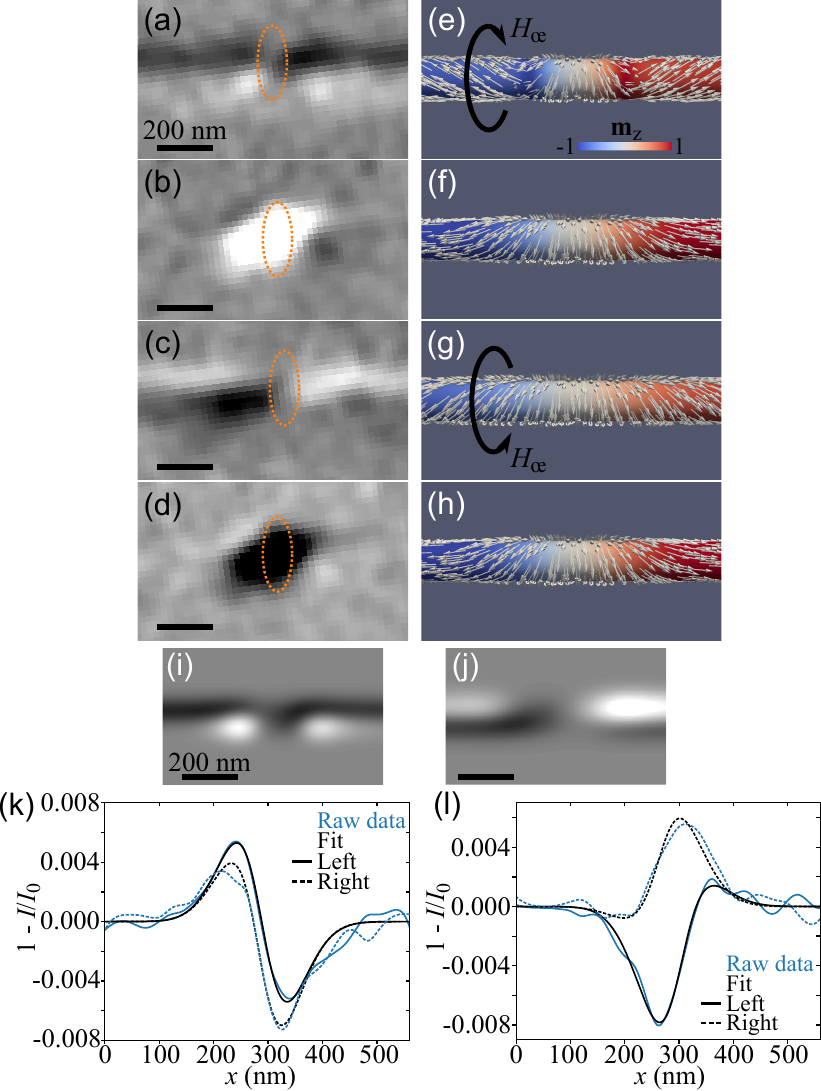}
\caption{BPW in a \SI{101}{nm} diameter NW subjected to an \OE rsted field. Average of all wavelet filtered time-resolved differential magnetic contrast frames acquired during (a) the application of a \SI{3}{ns} $(+)$ current pulse with amplitude $\SI{7.5\times{10^{11}}}{A/m^2}$, (b) the \SI{100}{ns} rest period, (c) the \SI{3}{ns} $(-)$ current pulse and (d) the \SI{100}{ns} rest period. The BPW position is indicated by orange circles. The micromagnetic simulations in (e-h) correspond to the images in (a-d) and show a tail-to-tail BPW in its compressed (a), expanded (g) and relaxed (f,h) state. The direction of the applied \OE rsted field is indicated where applicable. Simulated time-resolved differential magnetic contrast of a compressed (i) and expanded (j) BPW in a \SI{100}{nm} diameter NW. Each frame was blurred with a Gaussian. Contrast profiles for the case of a compressed (k) and expanded (l) BPW, corresponding to line scans left (solid lines) and right (dashed lines) of the wall center. The true line scans from the image in (a) and (c) are shown in blue, while the fit from our model is shown in black.}
\label{fig:TR-BPW}
\end{figure}

We now turn to the imaging of BPWs under the influence of the \OE rsted field in the same NWs. DW positions were determined using static XMCD, after which a current pulse of sufficient amplitude ($j\approx\SI{1\times{10^{12}}}{A/m^2}$) was sent through the NW to ensure a) the BPW DW type by transforming walls of the transverse-vortex kind to BPWs\cite{bib-FRU2021} and b) that the BPW is sufficiently pinned on an extrinsic pinning site to allow for a reproducible magnetization process over millions of pulses. Multiple time-resolved series were acquired while applying a similar voltage pulse signal as in \subfigref{fig:TRframes}{b}, with different $j$ around $j_c$, the critical current density expected for BPW circulation switching\cite{bib-FRU2021,bib-FRU2019b}. 

We imaged a tail-to-tail BPW in a \SI{101}{nm} diameter NW, first in the regime when $j<j_c$ by applying a current density of $\pm\SI{7.5\times{10^{11}}}{A/m^2}$. \subfigref{fig:TR-BPW}{a-d} show the frame average of the differential magnetic contrast frames acquired during (a) the application of the \SI{3}{ns} $(+)$ current pulse, (b) the first \SI{100}{ns} rest period, (c) the \SI{3}{ns} $(-)$ current pulse and (d) the second \SI{100}{ns} rest period. All frames were filtered with a hat wavelet filter to improve the signal-to-noise ratio. In addition to the \OE rsted field tilting observed within the domains during the pulse application, all four images contain a strong contrast at the BPW location (orange circle). 

As expected for this regime, in \subfigref{fig:TR-BPW}{a,c} no specific differential magnetic contrast is visible at the wall center, indicating that the sign of azimuthal curling directly around the Bloch-point remains unchanged for the duration of the image series. 
The applied \OE rsted field pulses (see black arrows in \subfigref{fig:TR-BPW}{e} and f) are thus once parallel and once antiparallel to the wall circulation, so that different dynamics are expected during each of the two pulses. Indeed, in (a) and (c) we observe two different stronger contrasts around the wall center.
In (a) there are four small symmetric lobes of differential magnetic contrast with a stronger intensity than seen in the longitudinal domains far from the wall. The bipolar contrast to either side of the wall center is indicative of a change of circulation that is now opposed to the static circulation (\ie state $I_\mathrm{S}$), which can be understood as follows: 
the bipolar contrast, albeit stronger, matches that within the domains, suggesting a tilt towards the \OE rsted field direction. This tilt direction opposes the intrinsic BPW circulation and as the BPW does not reverse its sign of circulation, the observed change therefore reflects rather a compression of the wall. Such a compression was predicted by micromagnetic simulations\bracketsubfigref{fig:TR-BPW}{e} of a tail-to-tail BPW in a \SI{100}{nm} diameter \CoNi{30}{70} NW, subjected to an antiparallel \OE rsted field induced by a current with amplitude $\SI{8\times{10^{11}}}{A/m^2}$. These simulations were obtained with our home-made finite element freeware \textsc{FeeLLGood}\cite{bib-FEE}, based on the Landau-Lifshitz-Gilbert equation. The simulated distribution of magnetization was further used to simulate XAS images\cite{bib-FRU2015c} expected for an absorption according to eq.~\eqref{eq:TR-intensity} and governed by our experimentally-determined values of  $\mu_{\mathrm{av}}$ and $\Delta\mu$. Importantly, the $\SI{30}{\degree}$ sample holder alignment was accounted for in the simulated imaging. Differential magnetic contrast images were calculated by applying eq.~\eqref{eq:TheoryTR} to simulated XAS images of a dynamic and static BPW. A Gaussian filter was applied to reproduce the effect of a finite width spot size as in the experiment (unfiltered images are shown in the supplementary material). The differential magnetic contrast simulated for a compressed tail-to-tail BPW is shown in \subfigref{fig:TR-BPW}{i}. The similarity with the image in (a) is striking, thus confirming the qualitative explanation of a BPW compression. Minor differences, \eg the size of certain features compared to in (a), are likely related to the Gaussian filter applied to the simulation.

Conversely, in (c) there are only two large lobes of opposite differential magnetic contrast, suggesting that the circulation of the static state is now being enhanced. This should be the result of an expansion of the BPW (see simulation in \subfigref{fig:TR-BPW}{g}) as the applied \OE rsted field is parallel to the wall circulation. The differential magnetic contrast from the simulated expanded wall (g) is shown in \subfigref{fig:TR-BPW}{j}, with key contrast features again matching with those observed in (c). This combination between time-resolved imaging and simulated imaging provides a powerful tool to explain observed contrasts.

We now return to our model for a quantitative verification of the qualitative explanation. We plot in blue in \subfigref{fig:TR-BPW}{k,l} contrast profiles from line scans taken across the wire, through the regions of contrast left (solid curve) and right (dashed curve) of the BPW center. \subfigref{fig:TR-BPW}{k} and (l) correspond to the images in (a) and (c), respectively. 
In this case, fitting is achieved because the static ($I_\mathrm{S}$) and the dynamic, either compressed or expanded, state intensity profiles are defined by separate $\theta_{0}$ and are non-zero. It must be noted that the convolution with the Gaussian spot size along $x$ only is now slightly less valid because the magnetization is no longer homogeneous along $z$, and a finer analysis should convolute along both the $x$ and $z$ direction. 
Still, the black curves in \subfigref{fig:TR-BPW}{k,l} are fits to the contrast profiles and show an excellent agreement on either side of the wall center. The BPW compression proposed to explain (a) is confirmed numerically with the fits in (k): left and right of the wall center, $\theta_{0}$ tilts from $16$ to \SI{-17}{\degree} and from $28$ to \SI{-8}{\degree}, respectively. The \OE rsted field reverses the sign of circulation close to the wall center, compressing the wall and giving rise to the bipolar contrast and hence four contrast lobes around the wall center. Similarly, the fits for (l) reveal $\theta_0$ tilts from $17$ to \SI{45}{\degree} and from $20$ to \SI{41}{\degree}, left and right of the wall center, respectively, confirming the enhancement of the static circulation, or an expansion of the BPW. 
(a) and (c) together show the breathing of the BPW, predicted only by simulations until now\cite{bib-FRU2021}. The differential magnetic contrast patterns were observed in multiple image series and are inverted in the case of a BPW with opposite static circulation (see supplementary material).

For the interpulse periods displayed in \subfigref{fig:TR-BPW}{b,d} a strong white or black contrast, respectively, is observed at the DW location. This should be a result of small scale BPW motion of the order of $270\pm\SI{30}{nm}$, however, intricacies of this contrast, such as the direction of motion and a time evolution of the contrast are not yet understood (see supplementary material for further discussion).

We finally mention the case when $j>j_c$, for which the BPW switches its sense of circulation\cite{bib-FRU2019b,bib-FRU2021}. This could not be observed experimentally with time-resolved STXM because DWs disappeared for $j	\geq j_c$, which we attribute to heat assisted DW depinning. Future measurements, possibly with engineered DW pinning sites are required to observe this effect with temporal resolution.



In conclusion, we have used time-resolved STXM to image dynamic changes of magnetic textures in cylindrical NWs. We observe the effect of the \OE rsted field on longitudinally magnetized domains and evidence the breathing of a BPW when subjected to pulses with opposite sign. A quantitative analysis of the differential magnetic contrast is provided by a robust model based on the absorptivity of X-rays and a description of the magnetization in a NW cross section. This highlights the depth of information obtainable with time-resolved magnetic imaging and that a direct comparison of the observed dynamics with simulations and theory is possible. Further work with this technique will significantly improve our understanding of magnetic 3D nanosized systems and enable better control over them.

See supplemental material for the following: full time-resolved image series from which still frames are shown in this text are shown in Fig.~S1 and Fig.~S2; calculation of the projection of magnetization in a tilted NW; detailed explanation of the XAS analysis, impact of the uncertainty in absorptivity on the asymmetric error bars on $\theta_0$; unfiltered simulated images of a breathing BPW; images of breathing of a BPW with switched circulation; and an explanation of differential magnetic contrast of BPW motion.

M. S. acknowledges a grant from the Laboratoire d’excellence LANEF in Grenoble (ANR-10-LABX-51-01). The project received financial support from the French National Research Agency (Grant No. JCJC MATEMAC-3D). This work was partly supported by the French RENATECH network, and by the Nanofab platform (Institut Néel), whose team is greatly acknowledged for
technical support. Part of the work was performed at the PolLux STXM endstation of the Swiss Light Source, Paul Scherrer Institut, Villigen PSI, Switzerland, financed by the German Bundesministerium für Bildung und Forschung (BMBF) through contracts 05K16WED and 05K19WE2.

The data that support the findings of this study are available from the corresponding author upon reasonable request.


\bibliographystyle{apsrev4-1}
\bibliography{Fruche8}

\end{document}


\title{Supplementary material: Time-resolved imaging of \OE rsted field induced magnetization dynamics in cylindrical magnetic nanowires}

\author{M.~Sch\"{o}bitz}
\email[Author to whom correspondence should be addressed: ]{michael.schobitz@cea.fr}
\affiliation{Univ.\ Grenoble Alpes, CNRS, CEA, Spintec, 38054 Grenoble, France}
\affiliation{Friedrich-Alexander Univ.\ Erlangen-N\"{u}rnberg, Inorganic Chemistry, 91058 Erlangen, Germany}
\affiliation{Univ.\ Grenoble Alpes, CNRS, Institut N\'{e}el, 38042 Grenoble, France}
\author{S.~Finizio}
\affiliation{Swiss Light Source, Paul Scherrer Institut, 5232 Villigen PSI, Switzerland}
\author{A.~De Riz}
\affiliation{Univ.\ Grenoble Alpes, CNRS, CEA, Spintec, 38054 Grenoble, France}
\author{J.~Hurst}
\affiliation{Univ.\ Grenoble Alpes, CNRS, CEA, Spintec, 38054 Grenoble, France}
\author{C.~Thirion}
\affiliation{Univ.\ Grenoble Alpes, CNRS, Institut N\'{e}el, 38042 Grenoble, France}
\author{D.~Gusakova}
\affiliation{Univ.\ Grenoble Alpes, CNRS, CEA, Spintec, 38054 Grenoble, France}
\author{J.-C.~Toussaint}
\affiliation{Univ.\ Grenoble Alpes, CNRS, Institut N\'{e}el, 38042 Grenoble, France}
\author{J.~Bachmann}
\affiliation{Friedrich-Alexander Univ.\ Erlangen-N\"{u}rnberg, Inorganic Chemistry, 91058 Erlangen, Germany}
\affiliation{Institute of Chemistry, Saint-Petersburg State Univ., 198504 St.\ Petersburg, Russia}
\author{J.~Raabe}
\affiliation{Swiss Light Source, Paul Scherrer Institut, 5232 Villigen PSI, Switzerland}
\author{O.~Fruchart}
\email[Author to whom correspondence should be addressed: ]{olivier.fruchart@cea.fr}
\affiliation{Univ.\ Grenoble Alpes, CNRS, CEA, Spintec, 38054 Grenoble, France}

\date{\today}

\maketitle

\section*{Full time resolved image series}
\begin{figure}
\centering
\includegraphics{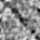}
\caption{Single frame showing the differential magnetic contrast observed in the time resolved image series D3\_{}43}
\label{fig:D3}
\end{figure}

The first image series\bracketsubfigref{fig:D3}{}  attached to this manuscript shows a full time-resolved image series from which still frames have been shown in the main manuscript (see Fig.~1c,d and Fig.~2a). The image series is comprised of 1021 frames, each taken at an interval of \SI{200}{ps} and a square pixel size of \SI{25}{nm}. A hat wavelet filter has been applied in order to reduce noise and improve the visibility of the magnetic contrast. In the image series, we observe the tilting of magnetization away from the nanowire (NW) long axis and towards the azimuthal \OE rsted field direction when a $(+)$ and $(-)$ polarity, \SI{3}{ns} current pulse with amplitude $\SI{1.1\times{10^{12}}}{A/m^2}$ is applied to a \SI{93}{nm} diameter \CoNi{30}{70} NW. We clearly see the effect of both of the pulses, as well as a very faint ringing effect that appears after each pulse and lasts for a duration of tens of ns.

\begin{figure}
\centering
\includegraphics{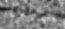}
\caption{Single frame showing the differential magnetic contrast observed in the time resolved image series C4\_{}86}
\label{fig:C4}
\end{figure}

The second image series\bracketsubfigref{fig:C4}{}  shows the full time-resolved series, from which still frames have been shown in the main manuscript in Fig.~3a-d. There are 1021 frames, taken at \SI{200}{ps} intervals and the same hat wavelet filtering was applied. In the image series, we observe the "breathing" of a Bloch-point domain wall (BPW) as explained in the main manuscript (see Fig.~3 of main manuscript), when a $(+)$ and $(-)$ polarity, \SI{3}{ns} current pulse with amplitude $\SI{7.5\times{10^{11}}}{A/m^2}$ is applied to a \SI{101}{nm} diameter \CoNi{30}{70} NW. In addition, in the interpulse time we can now visualize the slow growing and shrinking of a black and white contrast, respectively, indicating the slight motion of a BPW. This is further discussed later in this supplementary material.

\section*{Projection of magnetization in a tilted nanowire}
\begin{figure}
\centering
\includegraphics{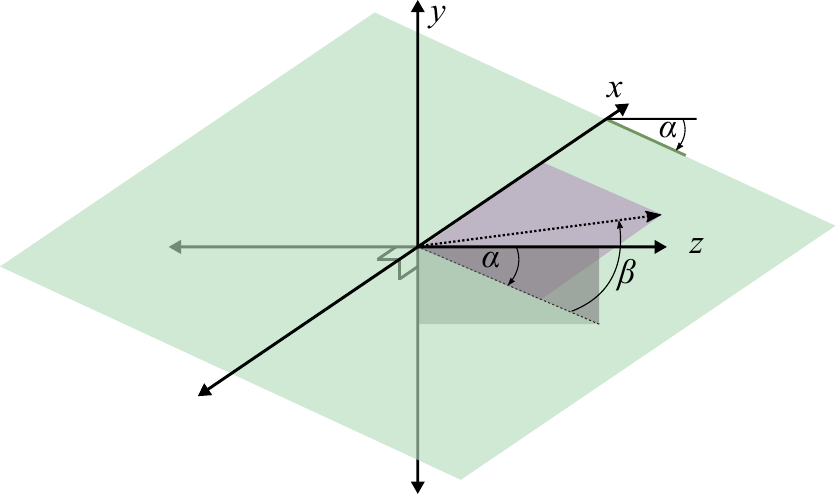}
\caption{Angles that must be considered to calculate the component of magnetization along the X-ray beam, the latter being in the $-y$ direction. The sample holder induces a tilt of angle $\alpha$, while the sample may be rotated in this plane by an angle $\beta$.}
\label{fig:Alignments}
\end{figure}

In the manuscript we use an Ansatz to describe the tilt by an angle $\theta$ of magnetic moment, $\vect{m}$, towards the azimuthal direction when subjected to the \OE rsted field,
\begin{equation}
\theta \left(r\right) = \theta_0 \sin{\left(\frac{\pi}{2}\frac{r}{R}\right)}
\label{eq:Ansatz}
\end{equation}
$\vect{r}$ is any point within the circular wire cross section and $r = \abs{\vect{r}}$. Further, $R$ is the wire radius and $\theta_0 = \theta \left( r = R\right)$. We now reformulate $\vect{m}$, which in spherical coordinates is $\left(1,\theta,\phi\right)$ on axes that are $(\hat{\vect{r}},\hat{\boldsymbol{\varphi}},\hat{\vect{z}}  )$ (see schematic in Fig.~2a in the manuscript). This can be projected into a cartesian coordinate system using a standard conversion, giving $\vect{m} = \left( \sin{\left(\theta\right)}\cos{\left(\varphi\right)}, \sin{\left(\theta\right)}\sin{\left(\varphi\right)}, \cos{\left(\theta\right)}\right)$ with $\varphi=\arctan{\left(y/x\right)}$ the angle between the $x$-axis and $\vect{r}$ (see schematic in Fig.~2a in the manuscript), where $(x,y)$ is the cartesian coordinate of $\vect{r}$. Further, we need to exhibit the component of $\vect{m}$ along the X-ray beam direction, $\hat{\vect{k}}$, given by the dot product $\hat{\vect{k}}\cdot\vect{m}$. This is partly defined by the plane of the sample holder, which is tilted by $\alpha=\SI{30}{\degree}$ from the $z$-axis and passing through the $x$-axis. Further, if the sample is rotated in the plane of the substrate, another angle, $\beta$, opens up, causing a deviation from the perfect alignment of the nanowire long axis with the beam direction. Both of these angles are represented in \subfigref{fig:Alignments}.
It follows that 
\begin{equation}
\hat{\vect{k}}\cdot\vect{m} = -    \sin\left(\theta \right)\sin\left(\varphi \right)\sin\left(\alpha \right)\sin\left(\beta \right)     -   \sin{\left(\theta\right)}\cos{\left(\varphi\right)}\cos{\left(\alpha\right)}      -      \cos{\left(\theta\right)}\sin{\left( \alpha\right)}\cos\left(\beta \right)
\label{eq:kdotm}
\end{equation}
As stated in the manuscript, \eqref{eq:kdotm} is directly related to the magnetic contribution to the absorption of X-rays and an exact solution related exclusively to $\theta_0$ is found by substituting $\theta$, $\varphi$, $\alpha$ and $\beta$.

\section*{X-ray absorption spectroscopy analysis}
\begin{figure}
\centering
\includegraphics{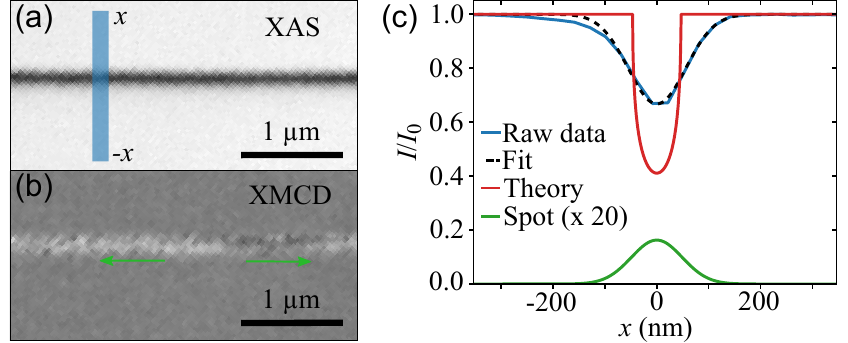}
\caption{Analysis of static XMCD STXM images to extract X-ray absorptivity and spot size. a) XAS image of a section of \SI{93}{nm} diameter \CoNi{30}{70} NW taken with static STXM using negative circularly polarized X-ray light at the Co L3 edge. In (b) the corresponding XMCD image of the same NW section shows two longitudinal magnetic domains, with their magnetization direction indicated (green). The intensity profile on the XAS image in (a) is plotted in (c) (blue). Also plotted are the fitted convoluted intensity profile (black), the theoretical transmitted intensity profile before convolution (red) and the Gaussian X-ray spot size (green, increased by factor 20).}
\label{fig:XAS-fit}
\end{figure}

The absorptivity of the studied \CoNi{30}{70} NWs at the Co L3 absorption edge is extracted from intensity profiles taken perpendicular to the wire long axis in X-ray absorption spectroscopy (XAS) images taken in static scanning transmission X-ray microscopy (STXM). A typical XAS image is shown in \subfigref{fig:XAS-fit}{a}. The line in blue indicates the line scan with which the intensity profile in blue in \subfigref{fig:XAS-fit}{c} is determined.
The observed profile is related to the absorptivity of X-rays and can be fitted with the Beer-Lambert law, according to which the theoretical transmitted intensity of X-ray light, $I\left(x\right)$, for a profile along $x$, is
\begin{equation}
I\left(x\right) = I_0 \exp \left[-\mu t\left(x\right)\right]
\label{eq:XAS-intensity}
\end{equation}
For the case of NWs, the thickness is $t\left(x\right) = 2 \sqrt{F \left( R^2-x^2\right)}$, where the $F=4/3$ term is a geometrical adjustment made for the $\SI{30}{\degree}$ sample holder and $R$ is the NW radius measured in SEM.
However, the intensity profile that can be calculated with \eqref{eq:XAS-intensity} assumes an infinitely small X-ray spot size. We model the realistic spot size with a Gaussian, normalized to have an integral equal to unity,
\begin{equation}
I_\mathrm{spot}\left(x\right) = \exp \left(-  x^2 / 2 \sigma^2  \right)  
\label{eq:Spot}
\end{equation}
where $\sigma$ is the standard deviation of the Gaussian, which is a measure of the spot width. We fit the observed XAS intensity profile with the convolution of \eqref{eq:XAS-intensity} and \eqref{eq:Spot} to accurately extract the only free parameters, $\mu$ and $\sigma$. The effect of the convolution is to spread the absorption profile laterally. This at the same time blurs the image and also decreases the apparent absorption, however, the total absorption remains fixed. This step is crucial to extract realistic values of $\mu$ with such an analysis.
The intensity profiles from the XAS images are nearly perfectly reproduced with the convoluted Beer-Lambert law (black dashed curve in \subfigref{fig:XAS-fit}{c}). Minor mismatches arise from the asymmetry in the true spot shape, however, these have little impact on the extracted values. The red and green curve in \subfigref{fig:XAS-fit}{c} display the transmitted X-ray intensity profile before convolution and the Gaussian spot size (scaled by a factor 20) used for said convolution, respectively.

However, due to the NW and sample holder orientation, the $\mu$ extracted by this analysis does not equal exactly the theoretical $\mu_{\pm,th}$ and we extract instead $\mu_\pm \approx \mu_{\mathrm{av}} \pm\frac{1}{2}\Delta\mu \left( \hat{\vect{k}}\cdot\vect{m}\right)$, which critically depends on the projection of magnetization along the X-ray beam direction. It is thus critical that we know the value of $ \hat{\vect{k}}\cdot\vect{m}$ for our analysis. 
To this end, the intensity profiles across the NWs (see blue line in \subfigref{fig:XAS-fit}{a}) were chosen in an area of uniform magnetization, \ie inside a longitudinally magnetized domain. This was known using the corresponding static XMCD images (see \subfigref{fig:XAS-fit}{b}) where homogeneous black or white contrasts indicated domains with uniform magnetization parallel or antiparallel to the wire long axis, respectively. 
Thus, due to the $\SI{30}{\degree}$ sample holder, $\hat{\vect{k}}\cdot\vect{m} = \pm \sin{\left(30\degree\right)} = \pm 0.5$ and as a result, the $\mu$ from our fit $\approx \mu_{\mathrm{av}} \pm\frac{1}{4}\Delta\mu$. Nonetheless, $\mu_{\mathrm{av}}$ and $\Delta\mu$ can be correctly calculated by carrying out this analysis on intensity profiles on multiple XAS images taken with both polarizations of light for the different NW samples and accounting for a factor 2 in $\Delta\mu$.

\section*{Asymmetric errors on $\theta_0$}
\begin{figure}
\centering
\includegraphics{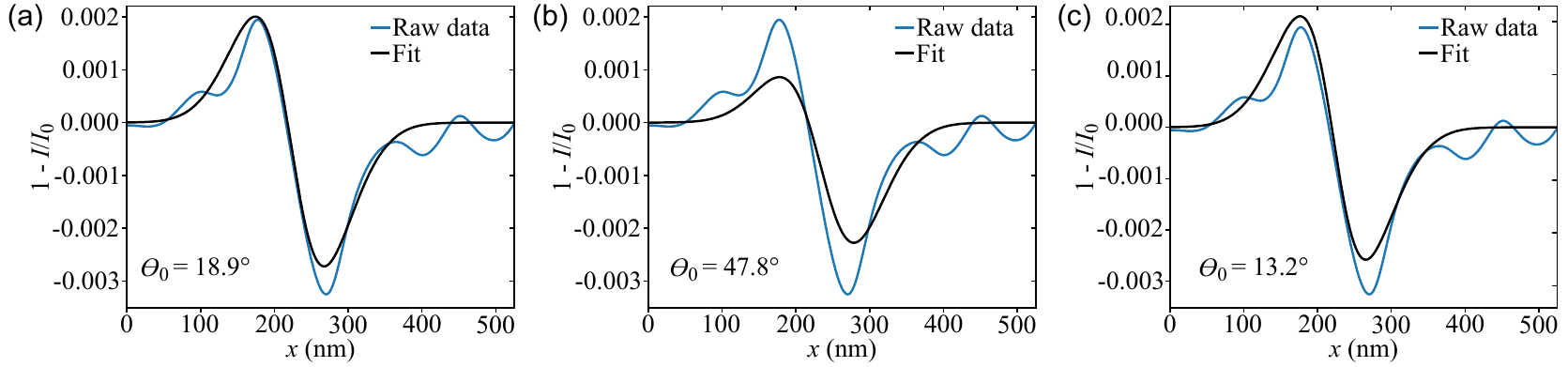}
\caption{Fitting magnetic contrast with different values of $\mu$. The raw data (blue) is fit by the model (black) presented in the main manuscript. In (a) the ideal parameters of $\Delta\mu=\SI{0.0032}{nm^{-1}}$ and $\mu_\mathrm{av}=\SI{0.011}{nm^{-1}}$ as noted in the main manuscript were used to giving an angle, $\theta_0 = \SI{18.9}{\degree}$. In (b) $\Delta\mu$ and $\mu_\mathrm{av}$ were increased by their one sigma uncertainty and the resulting angle is $\SI{47.8}{\degree}$. In (c) $\Delta\mu$ and $\mu_\mathrm{av}$ were decreased by their one sigma uncertainty, giving  $\theta_0 = \SI{13.2}{\degree}$.}
\label{fig:FittingError}
\end{figure}

As discussed in the manuscript, the uncertainty in the values of $\mu$ extracted in our XAS analysis leads to asymmetric error bars in $\theta_0$, due to the exponential nature of the Beer-Lambert law. However, this asymmetry is also reflected in the goodness of fit. In \subfigref{fig:FittingError}{a}, the raw data (blue) is fit by the model presented in the main manuscript (black) using the parameters of $\Delta\mu=\SI{0.0032}{nm^{-1}}$ and $\mu_\mathrm{av}=\SI{0.011}{nm^{-1}}$ extracted from the XAS analysis, giving an angle, $\theta_0 = \SI{18.9}{\degree}$. Further, the $r^2$ for this fit is $0.94$. In (b) the fit is repeated on the same raw data using instead $\Delta\mu=\SI{0.001}{nm^{-1}}$ and $\mu_\mathrm{av}=\SI{0.006}{nm^{-1}}$, which are the values extracted by the XAS analysis, but increased by one standard deviation. The resulting angle is $\SI{47.8}{\degree}$ and $r^2=0.83$. Finally in (c) the fit is repeated with $\Delta\mu=\SI{0.004}{nm^{-1}}$ and $\mu_\mathrm{av}=\SI{0.016}{nm^{-1}}$, \ie the values increased by one standard deviation, giving  $\theta_0 = \SI{13.2}{\degree}$ and $r^2 = 0.92$.
We see a significant difference in the goodness of fit between the upper\bracketsubfigref{fig:FittingError}{b} and lower error bound\bracketsubfigref{fig:FittingError}{c}. It should therefore be noted that while the uncertainty on the values of $\mu$ is appropriate, the upper bound for $\theta_0$ is normally an overestimation of the error. This is true for all of the results presented in Fig.~2d in the main manuscript.

\section*{Unfiltered simulated images of a breathing BPW}
\begin{figure}
\centering
\includegraphics{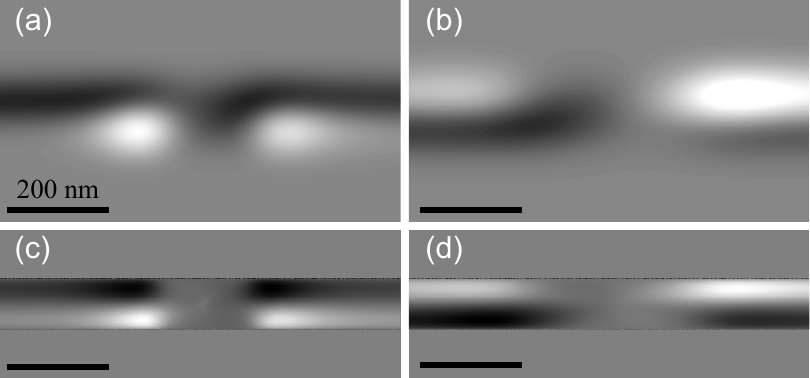}
\caption{Simulated time-resolved magnetic contrast of a compressed (a,c) and expanded (b,d) BPW in a \SI{100}{nm} diameter NW. The frames (a,b) have been filtered with a Gaussian in order to simulate a finite width spot size. The unfiltered frames are shown in (c,d).}
\label{fig:Simulation}
\end{figure}

The simulated magnetic contrast shown in Fig.~3i,j in the main manuscript are produced with a zero width spot size. In order to reproduce the magnetic contrast observed in the experimental images in Fig.~3a,c of the main manuscript, a Gaussian blur was applied to simulate a finite width spot size in the simulation. The filtered (a,b) and unfiltered (c,d) simulated magnetic contrast images are shown in \subfigref{fig:Simulation}.

\section*{Breathing of a BPW before and after intrinsic circulation switching}
\begin{figure}
\centering
\includegraphics{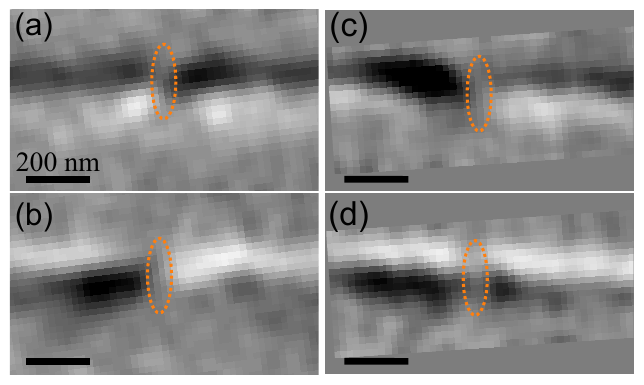}
\caption{Frame averages of two different time-resolved image series. The average of all frames acquired during the application of the \SI{3}{ns} duration $(+)$ current pulse is given in (a,c) while the average of all frames acquired during the application of the \SI{3}{ns} duration $(-)$ current pulse is given in (c,d). Panels (a,b) are reproduced from Fig.~4a,c of the main manuscript. Panels (c,d) are from a time-resolved image series acquired after an unexpected switching of circulation of the same BPW. The BPW center is indicated by an orange circle.}
\label{fig:Breathing}
\end{figure}

As mentioned in the main manuscript, the BPW breathing was observed in multiple time-resolved image series. In particular, the breathing of a wall was imaged before and after the wall unexpectedly reversed the sign of its intrinsic circulation in between two image series. As a result the contrast observed for the compression (and expansion) occurs during the opposite polarity pulse in these two image series. Frame averages of these two different time-resolved image series are shown in \subfigref{fig:Breathing}. The average of all frames acquired during the application of the \SI{3}{ns} duration $(+)$ current pulse is given in (a,c) while the \SI{3}{ns} $(-)$ current pulse is given in (c,d). (a,b) are reproduced from Fig.~4a,c of the main manuscript. A compression of the BPW occurs in (a) while, an expansion occurs in (b). The wall then unexpectedly reversed its circulation before the next image acquisition and as a result, the contrast expected for a BPW compression is observed in (d), during the pulse which earlier caused a wall expansion. Similarly, the expansion is observed in (c). Importantly, the \OE rsted field induced domain tilting remains the same, thus the swapped contrast cannot be a result of a reversed order of pulse polarity. This is yet another confirmation that this breathing effect is related to the intrinsic azimuthal circulation of the BPW.

\section*{Magnetic contrast of a BPW motion event}
\begin{figure}
\centering
\includegraphics{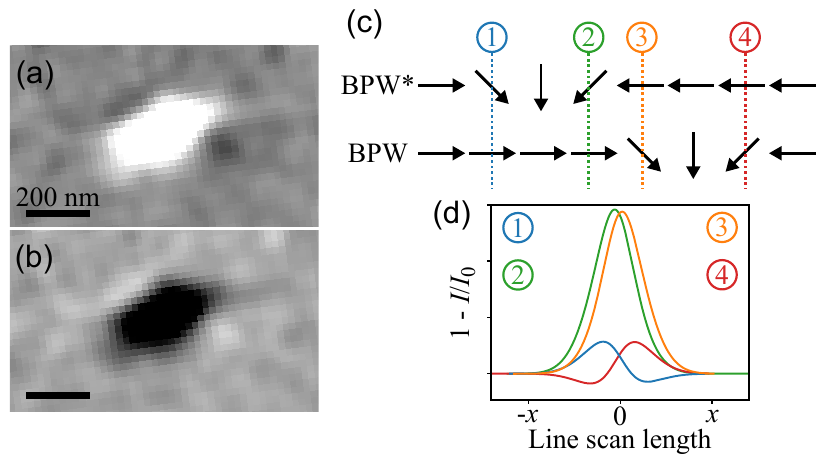}
\caption{Analysis of the magnetic contrast of BPW motion in a \SI{101}{nm} diameter NW. The average of all frames acquired during the interpulse periods after (a) the \SI{3}{ns} duration $(+)$ current pulse and (b) the \SI{3}{ns} duration $(-)$ current pulse, both with amplitude $\SI{7.5\times{10^{11}}}{A/m^2}$. Both (a,b) are shown as Fig.~4b,d in the main manuscript. c) Schematic of the arrangement of magnetic moments in a BPW and the same wall shifted slightly left (*), to perform an image calculation. The  dashed lines indicate the magnetic configuration used to predict the contrast profiles in (d).}
\label{fig:Motion}
\end{figure}

We note in the main manuscript that the strong white or black magnetic contrast observed in Fig.~4b,d (see also \subfigref{fig:Motion}{a,b}) is related to BPW motion. As the wall moves, the magnetization along the wire axis is entirely reversed compared to the state $I_\mathrm{S}$ and as such, a strong contrast is seen depending on the direction of motion. Particular, however, are the two dots of black (white) contrast spread diagonally around the strong white (black) contrast in \subfigref{fig:Motion}{a,b}. These are related to the intrinsic azimuthal circulation of the BPW and are expected where the wall motion begins and ends (\ie the left and right extremity of the strong contrast). As the wall moves, magnetization either tilts from the azimuthal to the longitudinal direction if the wall moves farther away, or conversely tilts from the longitudinal to the azimuthal direction if the wall moves closer. This tilting effect is phenomenologically similar to that due to an applied \OE rsted field, and as such a bipolar contrast should be expected across the wire. The bipolar contrast should be reversed to the left and right extremity of the motion event, because the tilt direction is opposite. We once more use our model to explain the contrast.
We first assume that the schematic in \subfigref{fig:Motion}{c} explains our wall motion event, where a 1D BPW initially located to the right is shifted to the left to become BPW*. Assuming the initial position provides the state $I_\mathrm{S}$, then the dynamic magnetic contrast profile expected at the four indicated locations (\#{1-4}) can be predicted with our model. Indeed, we find that to the left extremity of the motion event (position \#{1}) the contrast profile is asymmetric and bipolar as plotted in blue in \subfigref{fig:Motion}{d}. The contrast profile is very similar to that of the \OE rsted field tilting. In the center of the motion event (positions \#{2} and \#{3}) a strong uniform contrast is expected as plotted in green and orange in (d), which is explained by the complete reversal of magnetization along the wire axis. On the right extremity of the motion event (position \#{4}), the contrast profile is again asymmetric and bipolar, however with a reversed order (red curve in (d)). This describes rather well the contrast we observe in \subfigref{fig:Motion}{a,b} and lets us conclude that the contrast is a result of BPW motion.

However, when \subfigref{fig:Motion}{a,b} are viewed as individual frames, the contrast intensity varies with time (see attached image series C4\_{}86.tif). The physics behind this, whether an optical effect or magnetization dynamics, are not yet understood and remain an open question of this study.

\bibliographystyle{apsrev4-1}
\bibliography{Fruche8}